\documentclass{IEEEtran}
\usepackage{cite}
\usepackage{amsmath,amssymb,amsfonts}
\usepackage{graphicx}
\usepackage{textcomp,nicefrac}
\usepackage{here}
\def\BibTeX{{\rm B\kern-.05em{\sc i\kern-.025em b}\kern-.08em
T\kern-.1667em\lower.7ex\hbox{E}\kern-.125emX}}

\begin{document}
\title{Gigabit Ethernet Daisy-Chain on FPGA \\ for COMET Read-out Electronics}
\author{Eitaro Hamada, Yuki Fujii, Youichi Igarashi, Masahiro Ikeno, Satoshi Mihara, Hajime Nishiguchi, \\
Kou Oishi, Tomohisa Uchida, Kazuki Ueno, and Hiroshi Yamaguchi

\thanks{Eitaro Hamada, Youichi Igarashi, Masahiro Ikeno, Satoshi Mihara, Hajime Nishiguchi,
Tomohisa Uchida, Kazuki Ueno
are with Institute of Particle and Nuclear Studies, 
High Energy Accelerator Research Organization(KEK),
1-1 Oho, Tsukuba, Ibaraki 305-0801, Japan (e-mail: ehamada@post.kek.jp).}
\thanks{Yuki Fujii is with
School of Physics and Astronomy, Monash University, Clayton, Victoria 3800, Australia.}
\thanks{Kou Oishi is with
Department of Physics, Kyushu University, 744 Moto-oka, Nishi-ku, Fukuoka 819-0395, Japan}
\thanks{Hiroshi Yamaguchi is with
Applied Research Laboratory, High Energy Accelerator Research Organization(KEK), 1-1 Oho, 
Tsukuba, Ibaraki 305-0801, Japan}
\thanks{
This work was supported by the JSPS KAKENHI (Grant No. JP17H04841 and JP17H06135).}
}

\maketitle

\begin{abstract}
The COMET experiment at J-PARC aims to search for the neutrinoless transition of a muon to an electron.
We have developed the readout electronics board called ROESTI for the COMET straw tube tracker. We plan to install the ROESTI in the gas manifold of the detector. The number of vacuum feedthroughs needs to be reduced due to space constraints and cost limitations. In order to decrease the number of vacuum feedthroughs drastically, we developed a network processor with a daisy-chain function of Gigabit Ethernet for the FPGA on the ROESTI. 
We implemented two SiTCPs, which are hardware-based TCP processors for Gigabit Ethernet, in the network processor. 
We also added the data path controllers which handle the Ethernet frames and the event data.
The network processor enables ROESTI to process the slow control over UDP/IP and to transfer event data over TCP/IP. 
By using the network processor, we measured the throughput, the stability, and the data loss rate for two to six ROESTIs. 
In any number of boards, the throughput of the event data transfer achieved the theoretical limit of TCP over the Gigabit Ethernet stably and ROESTI stably sent 100\% of the data. 

\end{abstract}


\section{Introduction}
\label{sec:introduction}
\begin{figure}[t]
\centerline{\includegraphics[width=3.0in]{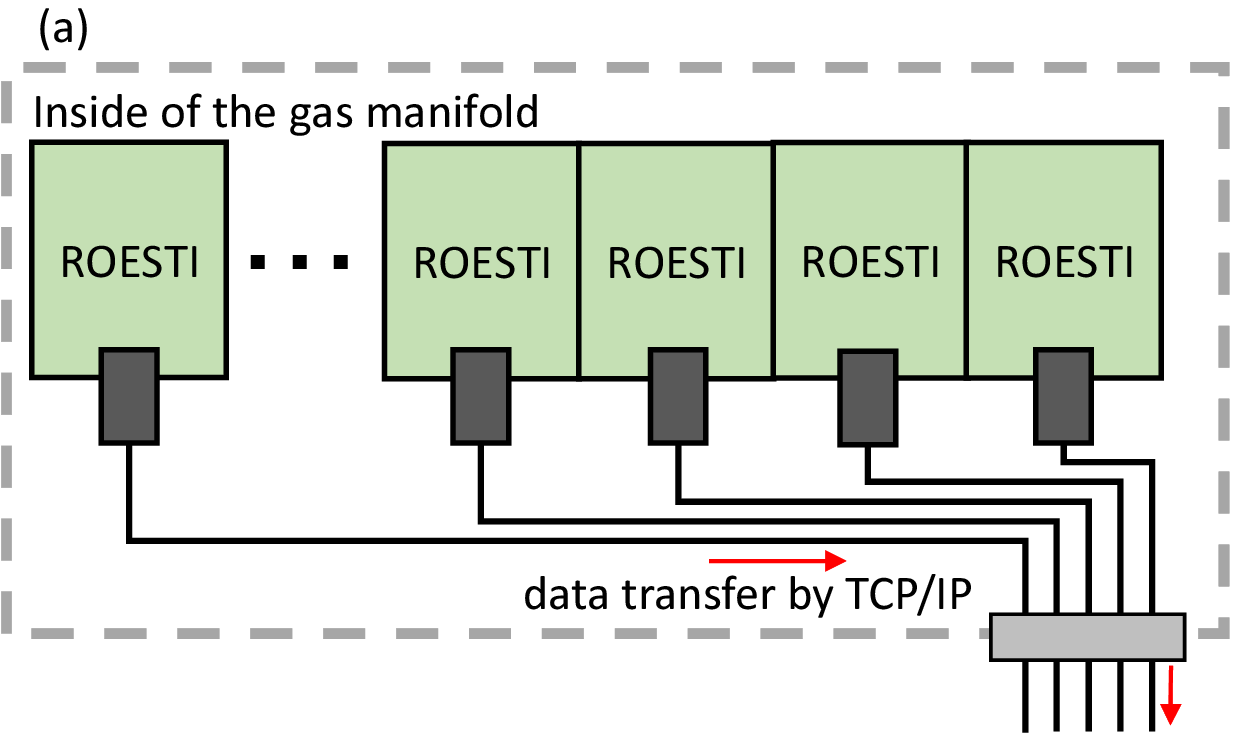}}
\centerline{\includegraphics[width=3.0in]{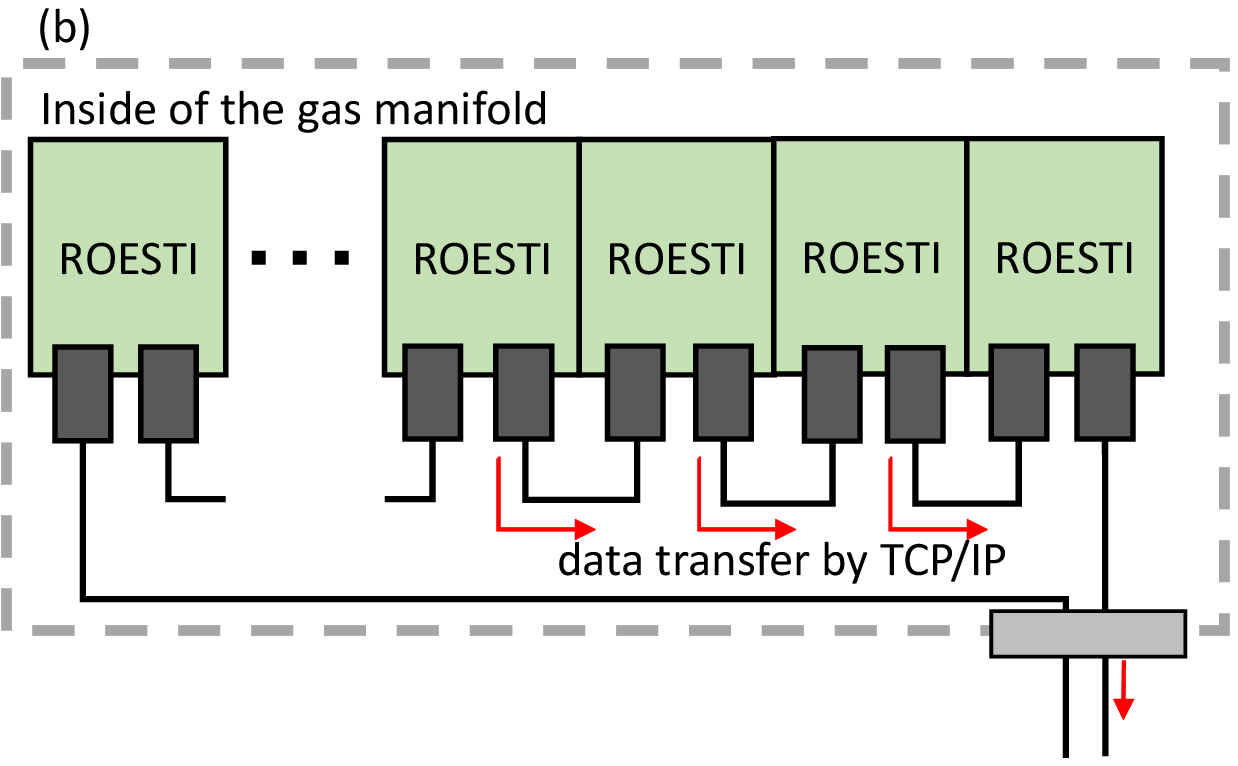}}
\caption{(a) Normal network topology (b) Network topology with the daisy-chain. The number of vacuum feedthroughs is decreased drastically.}
\label{fig:daisy}
\end{figure}

\IEEEPARstart{T}{he} 
COherent Muon to Electron Transition (COMET) experiment at J-PARC aims to search for the neutrinoless transition of a muon to an electron ($\mu$--$e$ conversion) in a muonic atom. Since charged lepton flavor is violated in this process, a branching ratio is highly suppressed to $\mathcal{O}(10^{-54})$ in the Standard Model (SM). However, theoretical models beyond the SM predict that the branching ratio of this process is to be $\mathcal{O}(10^{-15})$~\cite{ref:COMET}. Therefore, the discovery of $\mu$--$e$ conversion should be a clear evidence of new physics.
In order to suppress the background and to achieve the goal sensitivity, we adopt a straw tube tracker for the electron detector~\cite{ref:straw}. Since the detector is composed of an extremely light material which is operational in a vacuum, an excellent momentum resolution of better than 200 keV/c is achieved. We have developed the readout electronics board called ROESTI (Read-Out Electronics for Straw Tube Instrument) which reads out the signal from the detector precisely~\cite{ref:ROESTI1}. In order to prevent the degradation of the detector signal, the ROESTI needs to be located near the detector. We plan to install the ROESTI in the gas manifold of the detector. Data of the detector signal is transferred to the data acquisition (DAQ) PC with Ethernet~\cite{ref:ethernet}. This is because Ethernet provides advantages such as high-speed, high reliability, and high maintainability.
Although various kinds of Ethernet network topologies are created by using network switches, commercial network switches cannot be used in the gas manifold due to the space limitations and the radiation hardness. The heat of the ROESTI is cooled by the gas flow in the gas manifold.
Therefore, the number of the feedthroughs required for data communication lines should be the same as the one of the ROESTIs as shown in Fig.~\ref{fig:daisy}~(a).
However, we cannot adopt this network topology, because the number of vacuum feedthroughs needs to be reduced due to space constraints and cost limitations. 
To avoid this problem, we developed a daisy-chain function of Gigabit Ethernet for the FPGA on the ROESTI as shown in Fig.~\ref{fig:daisy}~(b). 
The required function for the communication with the daisy-chain is to transfer the data with TCP/IP~\cite{ref:TCP1} over Ethernet which provides high-speed and reliable communication. Additionally, the slow control function is required for setting and reading the parameters of the FPGA and IC chips on the ROESTI.
The performance target in terms of data transfer throughput on the daisy-chain is to achieve close to the maximum rate of TCP/IP over Gigabit Ethernet and to reduce data loss as much as possible. 
To enable the other ROESTIs to transfer data, the daisy-chain needs to transfer data bidirectionally as a fail-safe.

\begin{figure}[htb!]
 \begin{center}
 \includegraphics[width=0.45\textwidth]{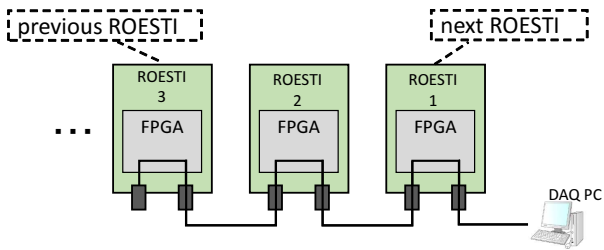}
 \end{center}
 \vspace{-5mm}
 \caption{Next ROESTI and previous ROESTI. The ROESTI1 is the next ROESTI from the ROESTI2. The ROESTI3 is the previous ROESTI from the ROESTI2.}
 \label{fig:sidedefine}
\end{figure}

Each ROESTI has two neighboring ROESTIs as shown in Fig.~\ref{fig:sidedefine}. 
We define the neighboring ROESTIs of near and far sides of the DAQ PC as "next ROESTI" and "previous ROESTI", respectively.

\section{ROESTI}
\label{sec:ROESTI}

\begin{figure}[htb!]
 \begin{center}
 \includegraphics[width=0.47\textwidth]{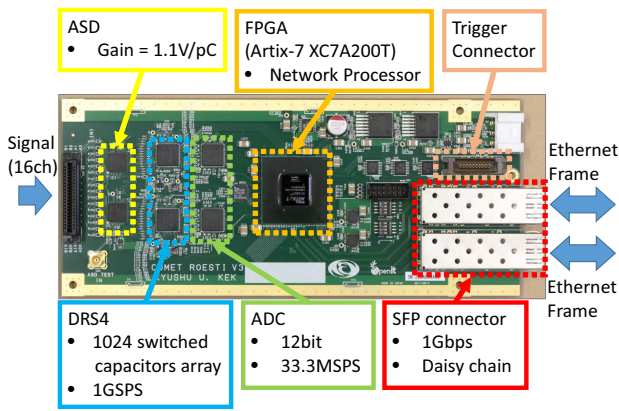}
 \end{center}
 \vspace{-5mm}
 \caption{Photograph of ROESTI prototype.}
 \label{fig:ROESTI}
\end{figure}

The ROESTI requires the gain of approximately 1\,V/pC and the signal-to-noise ratio (S/N) of larger than 5 at the minimum charge from the straw tube tracker. To achieve a momentum resolution of better than 200 keV/c on the tracker, a timing resolution of better than 1 ns is required. 
Event pileup capability is also needed due to the high hit rate.
Straw tube tracker has 2400 channels and ROESTI needs to read all of these channel signals.
Based on these requirements, we have developed a ROESTI prototype, which consists of 16-channel signal input connector, ASD~\cite{ref:ASD}, DRS4~\cite{ref:DRS4}, ADC (AD9637, Analog Devices), FPGA (XC7A200T-2FBG676C, Xilinx Inc.), Trigger connector, and Small Form-Factor Pluggable (SFP) connector. Fig.~\ref{fig:ROESTI} shows a photograph of the ROESTI prototype. The ASD amplifies and shapes the detector signal. The DRS4 and the ADC digitize analog signals with high-speed and high-accuracy. We evaluated the performance of this prototype and confirmed that it satisfies all the requirements described above~\cite{ref:ROESTI1}.

\begin{figure}[htb!]
 \begin{center}
 \includegraphics[width=0.45\textwidth]{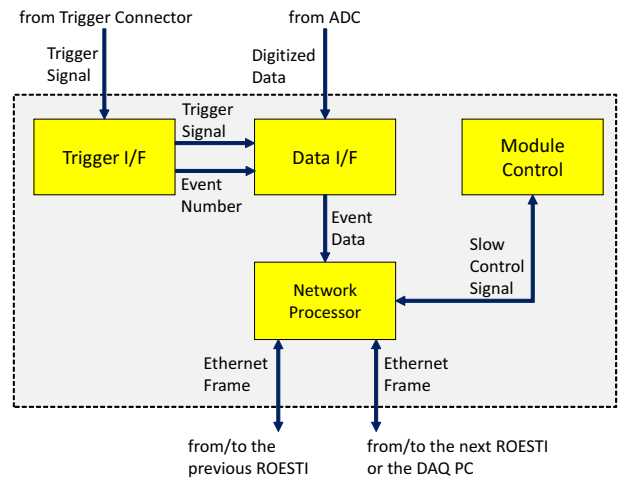}
 \end{center}
 \vspace{-5mm}
 \caption{Block diagram of FPGA.}
 \label{fig:FPGA}
\end{figure}

\begin{figure}[htb!]
 \begin{center}
 \includegraphics[width=0.35\textwidth]{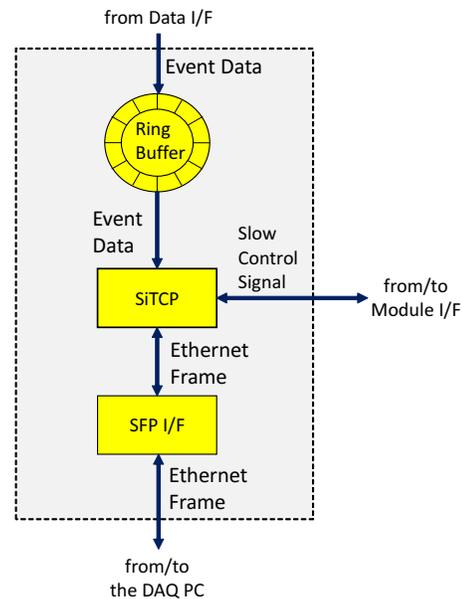}
 \end{center}
 \vspace{-5mm}
 \caption{Block diagram of Network Processor.}
 \label{fig:beforeNetwork}
\end{figure}

Fig.~\ref{fig:FPGA} shows a block diagram of the FPGA. The FPGA consists of \textit{Trigger I/F}, \textit{Data I/F}, \textit{Module Control}, and \textit{Network Processor}. After the FPGA receives the trigger signal from the trigger connector, the trigger signal is sent to the \textit{Trigger I/F}. In the \textit{Trigger I/F}, the trigger signal is handled and the event number is generated. Every time the trigger signal is received, the event number is incremented by one. The trigger signal and the event number are sent to \textit{Data I/F}.
After receiving the trigger signal, \textit{Data I/F} starts to receive digitized data from the ADC and to generate event data. The event data includes both the digitized data and the event number. The event data is sent to the \textit{Network Processor}.
Fig.~\ref{fig:beforeNetwork} shows a block diagram of the \textit{Network Processor}. The \textit{Network Processor} consists of ring buffer, SiTCP~\cite{ref:SiTCP}, and SFP I/F. After being temporarily stored in the ring buffer, the event data is sent to the SiTCP. The SiTCP is a hardware-based TCP processor for the Gigabit Ethernet and is designed for small devices limited by hardware size, such as an FPGA in front-end devices. 
The SiTCP enables a user circuit to process receiving and transmitting over TCP/IP.
Using the SiTCP, the event data are encapsulated in the TCP packet. 
The SiTCP provides a mechanism for slow control over UDP/IP~\cite{ref:UDP}.
In the slow control process with the SiTCP, at first, the DAQ PC must send a UDP packet for requests to the SiTCP. 
When the SiTCP in the \textit{Network Processor} receives the UDP packet, the SiTCP extracts the slow control signal from the UDP packet and sends the signal to the \textit{Module control}. 
The slow control signal includes address and data which are needed for the process of the slow control.
The \textit{Module control} receives the slow control signal and processes it. After completing the slow control process, the 
\textit{Module control} sends the slow control signal to SiTCP as an acknowledge signal.
The SiTCP encapsulates the slow control signal in the UDP packet and sends the UDP packet to the DAQ PC.
In order to communicate by TCP/IP or UDP/IP, the SiTCP encapsulates the TCP packet or the UDP packet in an Ethernet frame~\cite{ref:ethernet} which includes the destination MAC address. By exchanging the Ethernet frame with other network devices, the SiTCP can communicate over Ethernet. 
The SFP I/F provides physical layer process such as Physical Coding Sublayer (PCS) and Physical Medium Attachment (PMA). This module connects the Ethernet frame to the SFP connector and created using Xilinx 1G/2.5G BASE-X PCS/PMA Core~\cite{ref:IP}.

\section{DAISY-CHAIN IMPLEMENTATION IN FPGA}
\label{sec:FPGA}

\begin{figure}[htb!]
 \begin{center}
 \includegraphics[width=0.45\textwidth]{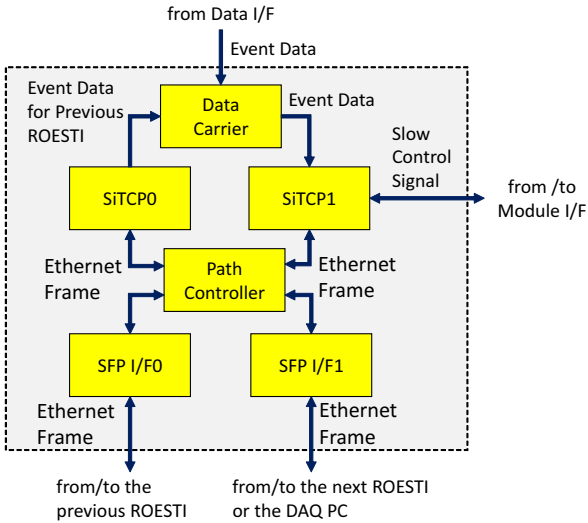}
 \end{center}
 \vspace{-5mm}
 \caption{Block diagram of new Network Processor.}
 \label{fig:blockdaisychain}
\end{figure}

We improved the \textit{Network Processor} and developed the daisy-chain function.
Another SiTCP and SFP I/F were added in the network processor.
Data Path controllers for the daisy-chain function were also implemented around them.
Fig.~\ref{fig:blockdaisychain} shows a block diagram of the new \textit{Network Processor}. This block consists of \textit{Path Controller}, \textit{Data Carrier}, and two SFP I/Fs (SFP I/F0 and SFP I/F1), along with the two SiTCPs (SiTCP0 and SiTCP1). 
The \textit{Path Controller} manages the Ethernet frame path.
The \textit{Data Carrier} controls the event data path.
SiTCP0 exchanges Ethernet frames with the previous ROESTI. SiTCP1 exchanges Ethernet frames with the next ROESTI or the DAQ PC.

\subsection{Path Controller}
\label{sec:pathcont}

\begin{figure}[htb!]
 \begin{center}
 \includegraphics[width=0.45\textwidth]{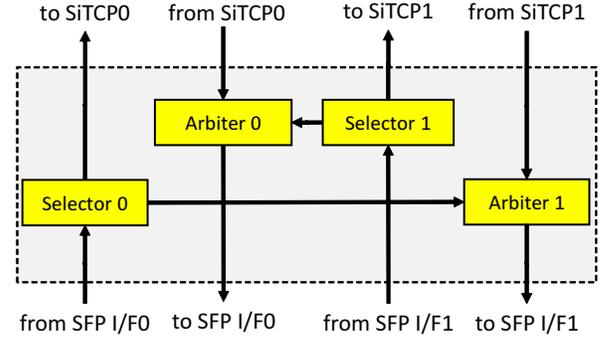}
 \end{center}
 \vspace{-5mm}
 \caption{Block diagram of Path Controller.}
 \label{fig:blockpathcont}
\end{figure}

The \textit{Path Controller} has the following two functions.
The first one is a receive function. Each ROESTI receives an Ethernet frame from the neighboring ROESTI. When the \textit{path controller} receives the Ethernet frame, the \textit{Path Controller} verifies the destination MAC address of the Ethernet frame and decides whether to send the Ethernet frame to own SiTCP or the other neighboring ROESTI. The second one is the send function. The \textit{Path Controller} receives the Ethernet frame from own SiTCP or the neighboring ROESTI. 
The \textit{Path Controller} arbitrates between two Ethernet frames and sends the Ethernet frame to the neighboring ROESTI or the DAQ PC. The Ethernet frames are synchronized with 125~MHz clock.

The \textit{Path Controller} consists of two \textit{Selector}s (\textit{Selector}0 and \textit{Selector}1) and two \textit{Arbiter}s (\textit{Arbiter}0 and \textit{Arbiter}1) as shown in Fig.~\ref{fig:blockpathcont}.
Each \textit{Selector} receives the Ethernet frame and verifies the destination MAC address thereof. If the destination MAC address matches the MAC address of the SiTCP, the \textit{Selector} sends the data to the SiTCP. If these MAC addresses are different, the \textit{Selector} sends the data to an \textit{Arbiter}. Alternatively, if the destination MAC address matches the broadcast address (for Ethernet, FF:FF:FF:FF:FF:FF), the \textit{Selector} sends the data to both the SiTCP and the \textit{Arbiter}. 
The \textit{Arbiter}0 can receive the Ethernet frame only from the \textit{Selector}1 or the SiTCP0. On the other hand, The \textit{Arbiter}1 can receive the Ethernet frame only from the \textit{Selector}0 or the SiTCP1.
When each \textit{Arbiter} receives the Ethernet frame, the \textit{Path Controller} starts transferring the Ethernet frame to the SFP I/F. If the \textit{Arbiter} receives other data during this process, the two Ethernet frames collide. In such cases, the \textit{Arbiter} discards the latter Ethernet frame.
For example, if the \textit{Arbiter}1 receives an Ethernet frame from the SiTCP1 when the \textit{Arbiter}1 does not send any data, the \textit{Arbiter}1 starts to transfer the Ethernet frame to SFP I/F1. During the transfer of the Ethernet frame, if the \textit{Arbiter}1 receives the other Ethernet frame from the \textit{Selector}0, the \textit{Arbiter}1 discards the Ethernet frame from the \textit{Selector}0.

\subsection{Data Carrier}
\label{sec:datacarrier}

\begin{figure}[htb!]
 \begin{center}
 \includegraphics[width=0.45\textwidth]{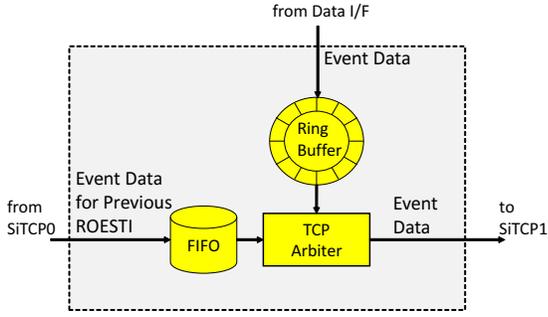}
 \end{center}
 \vspace{-5mm}
 \caption{Block diagram of Data Carrier.}
 \label{fig:blockdatacarieer}
\end{figure}

\begin{figure}[htb!]
 \begin{center}
 \includegraphics[width=0.33\textwidth]{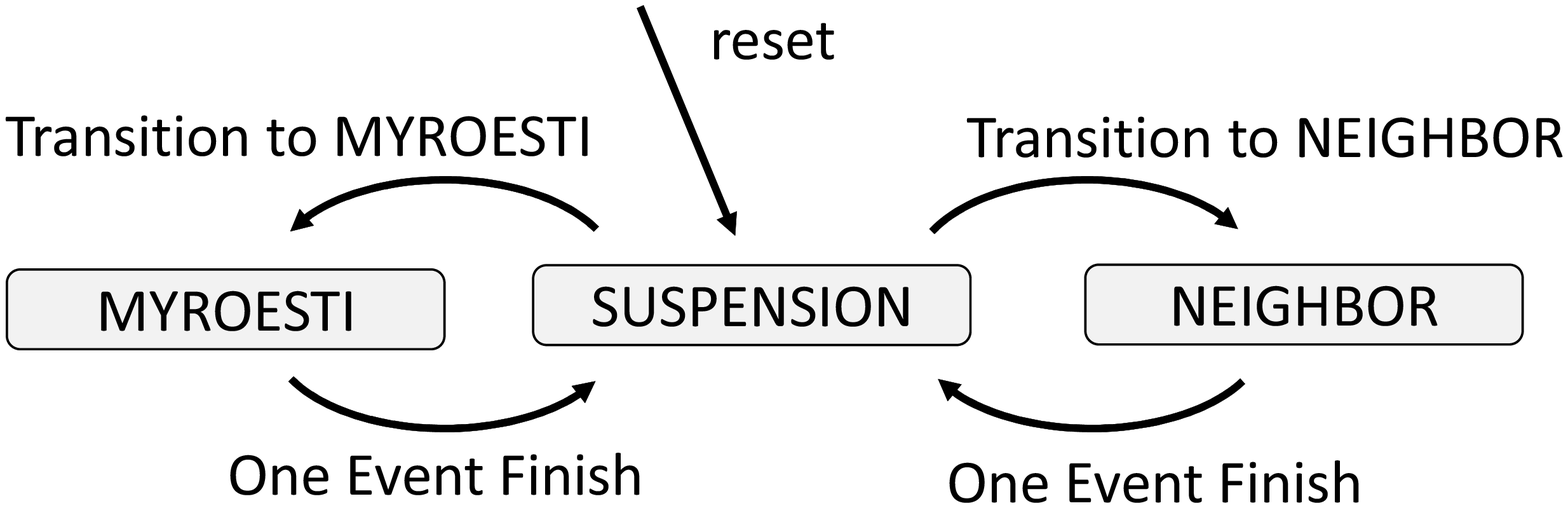}
 \end{center}
 \vspace{-5mm}
 \caption{State transition diagram for the TCP Arbiter of the Data Carrier.}
 \label{fig:State}
\end{figure}

In the \textit{Data Carrier}, the event data path is controlled. The \textit{Data Carrier} receives the event data from the \text{Data I/F}. The \textit{Data Carrier} also receives the event data for the previous ROESTI from SiTCP0. In order to avoid Ethernet frames collision in the Path Controller, Ethernet frames of TCP of previous ROESTI are sent to the Data Carrier. The \textit{Data Carrier} arbitrates between two event data and sends the event data to SiTCP1. The event data is synchronized with 133~MHz clock.

Fig.~\ref{fig:blockdatacarieer} shows the \textit{Data Carrier} block. The \textit{Data Carrier} consists of ring buffer, FIFO, and \textit{TCP Arbiter}.
The ring buffer stores the event data from the \text{Data I/F}. The FIFO stores the event data from SiTCP0. The ring buffer size is 64~bit $\times$ 4096~bit. The FIFO size is 8~bit $\times$ 65536~bit. 
The \textit{TCP Arbiter} extracts the event data from the ring buffer or the FIFO and sends the event data to SiTCP1.
The \textit{TCP Arbiter} has three states: SUSPENSION, MYROESTI, and NEIGHBOR as shown in Fig.~\ref{fig:State}. When the \textit{TCP Arbiter} receives the reset signal, the \textit{TCP Arbiter} moves to the SUSPENSION state. If the ring buffer has the event data for one or more events and the FIFO does not have any event data, the state changes to MYROESTI (Transition to MYROESTI). In the opposite case, the state changes to NEIGHBOR (Transition to NEIGHBOR). When both the ring buffer and the FIFO have the event data for one or more events, the state is determined by the event number of the event data in the ring buffer and the FIFO. If the event number of the event data in the ring buffer is less than that in the FIFO, the \textit{TCP Arbiter} state changes to MYROESTI (Transition to MYROESTI); otherwise, it changes to NEIGHBOR (Transition to NEIGHBOR). When the state changes to MYROESTI, the \textit{TCP Arbiter} extracts one event data from the ring buffer and sends it to SiTCP1. In contrast, when the state changes to NEIGHBOR, the \textit{TCP Arbiter} extracts one event data from the FIFO and sends it to SiTCP1. Upon transmitting the event data, the state returns to SUSPENSION (One Event Finish).

\subsection{Slow Control over UDP/IP}
\label{sec:slowcontrol}



\begin{figure}[htb!]
 \begin{center}
 \includegraphics[width=0.44\textwidth]{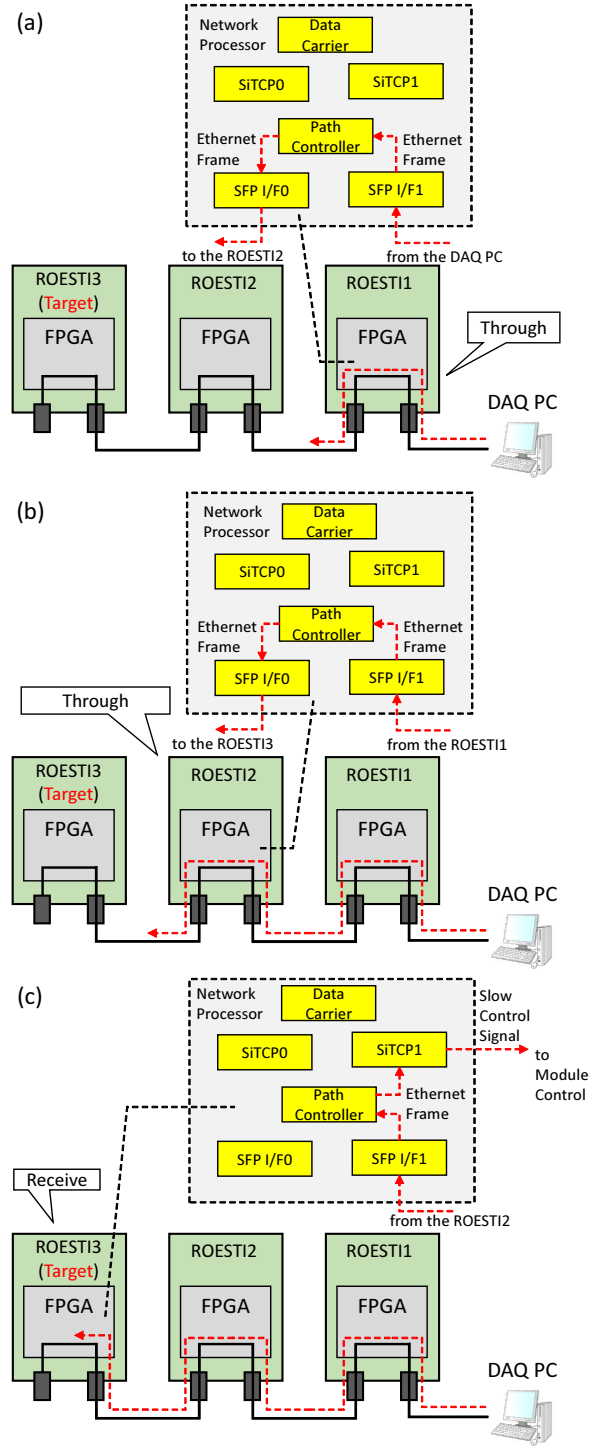}
 \end{center}
 \vspace{-5mm}
 \caption{Example of slow control over UDP/IP. (a) Ethernet frame passes through the ROESTI1 because of MAC address mismatch. (b) Ethernet frame also passes through the ROESTI2 because of MAC address mismatch. (c) Ethernet frame is received in the ROESTI3 because of MAC address match. The slow control signal is sent to the \textit{Module Control} of the ROESTI3. }
 \label{fig:egUDP}
\end{figure}

As mentioned in Section~\ref{sec:ROESTI}, the DAQ PC must send the slow control signal to SiTCP in the slow control process. After processing the slow control, SiTCP sends a slow control signal to the DAQ PC. For daisy-chain communication, the same process takes place between the DAQ PC and the target ROESTI. 

Fig.~\ref{fig:egUDP} shows an example of the slow control process over UDP/IP. The DAQ PC and three ROESTIs (ROESTI1, ROESTI2, ROESTI3) are connected with the daisy-chain. The ROESTI3 and the DAQ PC communicate for the slow control over UDP/IP. The DAQ PC sends the Ethernet frame such that the destination MAC address matches the MAC address of the SiTCP1 in the ROESTI3. 
The SiTCP1 of each ROESTI has a different MAC address. Because the MAC address of the SiTCP1 of the ROESIT1 does not match the destination MAC address of the Ethernet frame, the Ethernet frame simply passes through the \textit{Path Controller} of the ROESTI1 and is sent to the ROESTI2 (Fig.~\ref{fig:egUDP} (a)). For the same reason, the Ethernet frame passes through the ROESTI2 and is sent to the ROESTI3 (Fig.~\ref{fig:egUDP} (b)). 
Subsequently, the \textit{Path Controller} of the ROESTI3 identifies the Ethernet frame and sends it to the SiTCP1 of the ROESTI3 (Fig.~\ref{fig:egUDP} (c)). The Ethernet frame includes the slow control signal. The slow control signal is sent to the \textit{Module Control} of the ROESTI3. 
After that, the ROESTI3 sends the Ethernet frame to the DAQ PC. The destination MAC address corresponds with the MAC address of the DAQ PC. The Ethernet frame passes through the ROESTI1 and the ROESTI2 again before reaching its destination.
The slow control process does not change even if the number of ROESTIs increases. 

\begin{figure}[htb!]
 \begin{center}
 \includegraphics[width=0.5\textwidth]{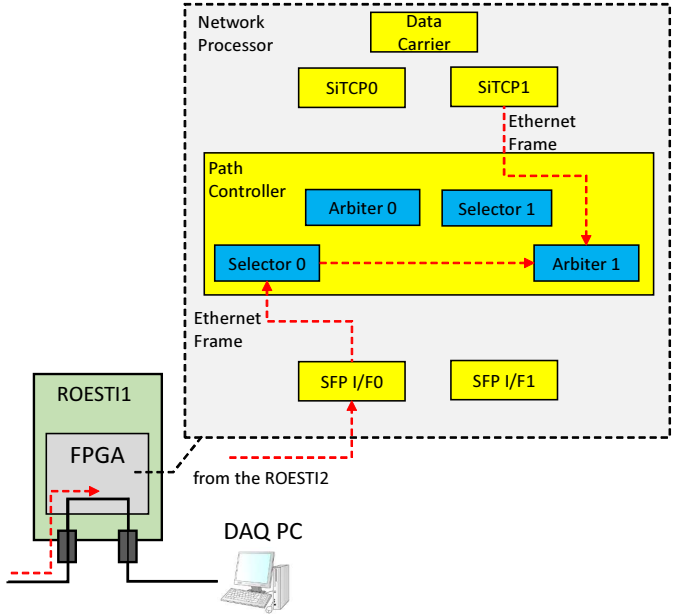}
 \end{center}
 \vspace{-5mm}
 \caption{Data collision in the Arbiter1 of the Path Controller. }
 \label{fig:data collision}
\end{figure}

In the \textit{Arbiter} of the \textit{Path Controller}, if the timing of two Ethernet frame matches as shown in Fig.~\ref{fig:data collision}, they collide with each other. Either of the Ethernet frames disappears as mentioned in Section~\ref{sec:pathcont}. However, this is not a major problem in the case of the slow control over UDP/IP because the frequency of slow control is small enough compared to the bandwidth of Gigabit Ethernet and slow control usually process when ROESTI does not transfer event data with TCP. 
If the slow control process when ROESTI transfers event data with TCP, It is possible to disappear Ethernet frame of slow control. However,  the DAQ PC finds this anomaly and the DAQ PC resends the Ethernet frame for slow control to the target ROESTI until slow control succeed. This is not a problem because the slow control process does not need instantaneousness.

\subsection{Data Transfer over TCP/IP}
\label{sec:datatransfer}

\begin{figure}[htb!]
 \begin{center}
 \includegraphics[width=0.5\textwidth]{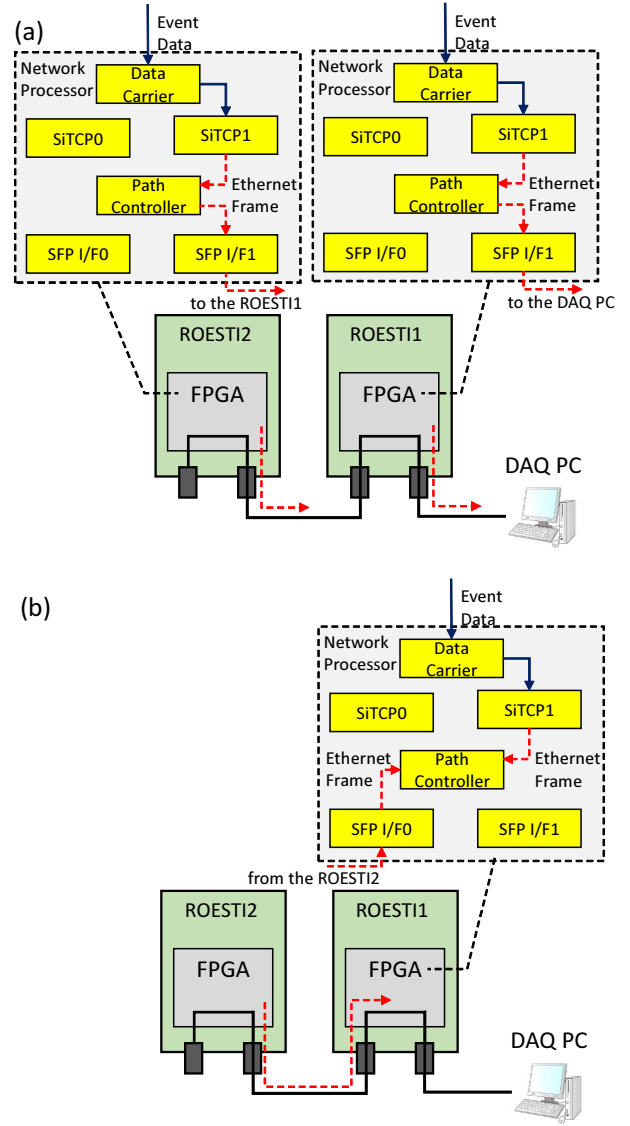}
 \end{center}
 \vspace{-5mm}
 \caption{Example 1 of transferring event data over TCP/IP. (a) Two ROESTIs receives the trigger signal and the event data is generated. (b) Two Ethernet frames collide in the Path Controller of the Network Processor. }
 \label{fig:egTCP1}
\end{figure}

Unlike the slow control process over UDP/IP, the ROESTIs and DAQ PC cannot communicate directly in the event data transfers over TCP. 
Fig.~\ref{fig:egTCP1} shows an example of the event data transfer process over TCP/IP. The DAQ PC and two ROESTIs (ROESTI1, ROESTI2) are connected with the daisy-chain. The ROESTI2 attempts to send the Ethernet frame to the DAQ PC directly. 
When each ROESTI receives the trigger signal, the event data is generated on each ROESTI (Fig.~\ref{fig:egTCP1}~(a)). The event data is sent to SiTCP1 of \textit{Network Processor} and encapsulated in the Ethernet frame. Each Ethernet frame is sent to the DAQ PC. 
The destination MAC address of the Ethernet frame from ROESTI2 corresponds with the MAC address of the DAQ PC. 
The Ethernet frame from the ROESTI2 attempts to pass through in the \textit{Path Controller} of the \textit{Network Processor} (Fig.~\ref{fig:egTCP1}~(b)).
If the Ethernet frame is sent to the \textit{Path Controller} from SiTCP1 at the same time, two Ethernet frames collide in the \textit{Path Controller} as shown in Fig.~\ref{fig:data collision}.
Either of the Ethernet frames disappears as mentioned in Section~\ref{sec:pathcont}. 
In such cases, the SiTCP1 of the ROESTI1 or the ROESTI2 detects the loss of the Ethernet frame. The SiTCP1 subsequently resends the same Ethernet frame. This process is called TCP re-transmission~\cite{ref:TCP1}. 
The TCP re-transmission might slow down the data transfer speed and cause unstable communication.
Therefore, this communication method cannot be used.

\begin{figure}[htb!]
 \begin{center}
 \includegraphics[width=0.5\textwidth]{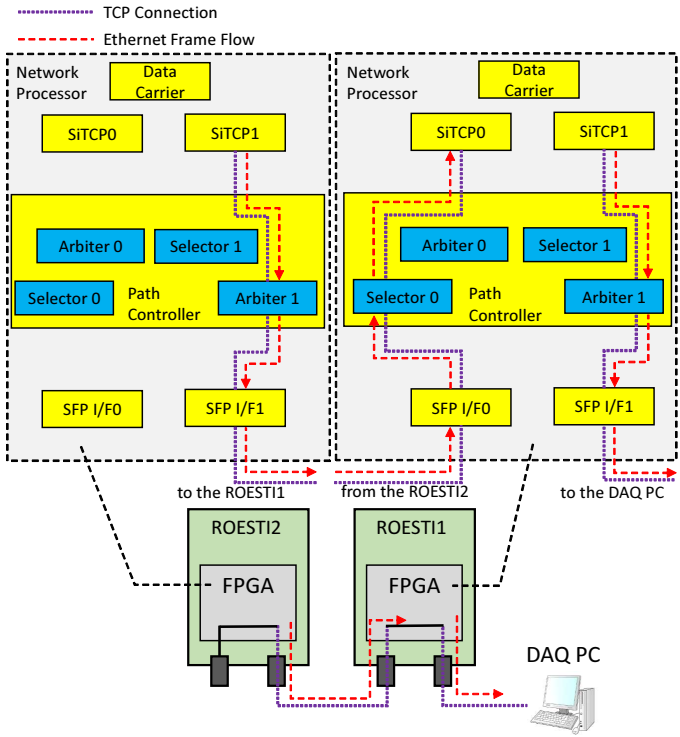}
 \end{center}
 \vspace{-5mm}
 \caption{No data collision in the Arbiter1 of the Path Controller.}
 \label{fig:nodatacollision}
\end{figure}

To avoid the TCP re-transmission, TCP connection is established between SiTCP1 of one ROESTI, which is the closest to the DAQ PC, and the DAQ PC as shown in Fig.~\ref{fig:nodatacollision}. This ROESTI sends the event data to the DAQ PC.
TCP connection is also established between SiTCP1 of another ROESTI and SiTCP0 of next ROESTI. These ROESTIs send the event data to the next ROESTI.
In such cases, the ROESTI receives the Ethernet frame, which the destination is own ROESTI, from the previous ROESTI. Because of MAC address match, the \textit{selector0} of the \textit{Path Controller} sends the Ethernet frame to SiCTP0 and the \textit{Arbiter} in the \textit{Path Controller} receives only one Ethernet frame as shown in Fig.~\ref{fig:nodatacollision}. Therefore, data collision does not occur in the \textit{Path Controller}.

\begin{figure*}[t]
 \begin{center}
  \includegraphics[width=0.74\textwidth]{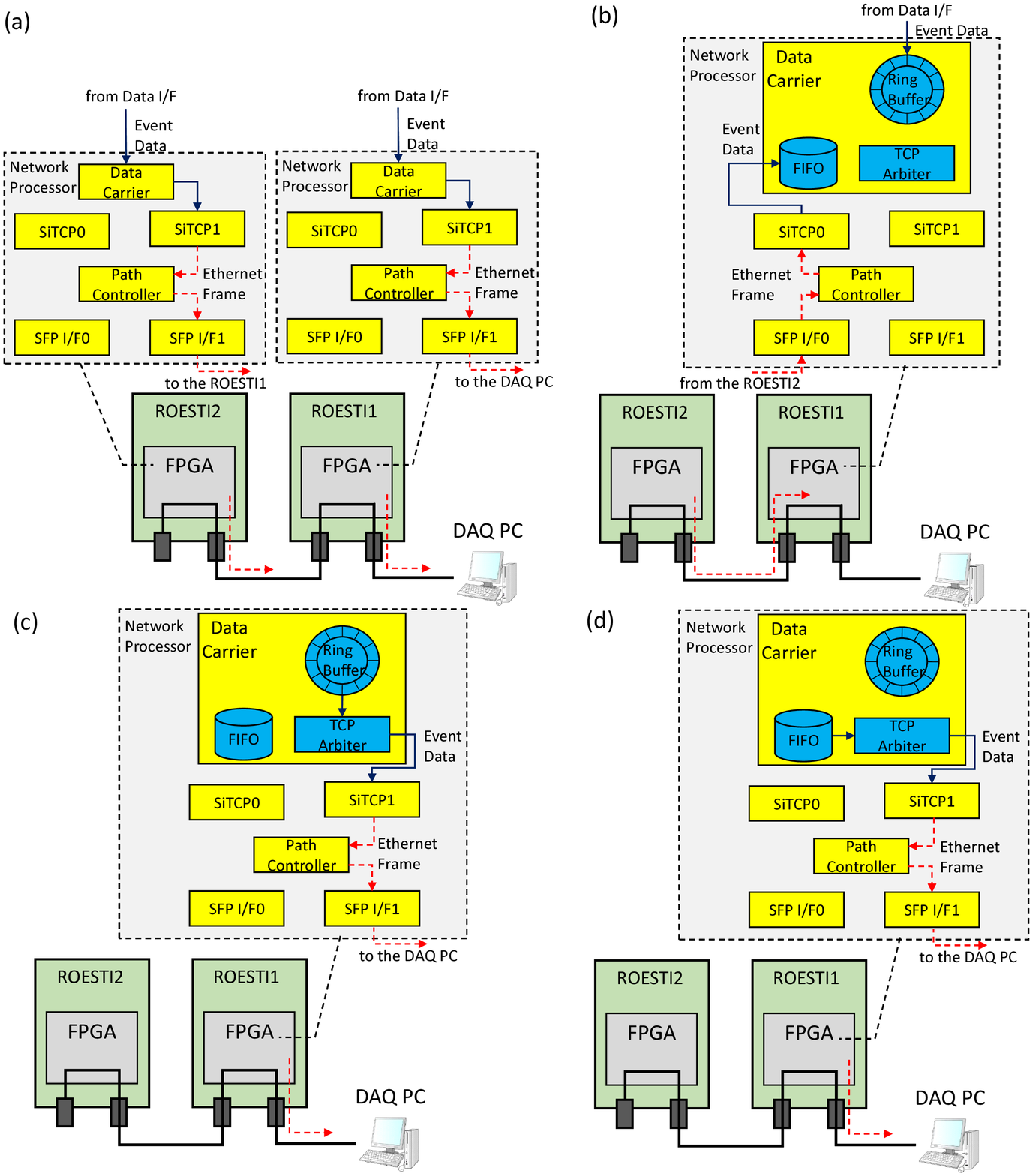}
  \caption{Example 2 of transferring event data over TCP/IP. (a) Two ROESTIs receive the trigger signal and the event data is generated in the Data I/F. (b) The Ethernet frame is sent to the ROESTI1 from the ROESTI2. (c) The event data in the ring buffer is sent to the DAQ PC. (d) The event data in the FIFO is sent to the DAQ PC. }
  \label{fig:egTCP2}
 \end{center}
\end{figure*}

Fig.~\ref{fig:egTCP2} shows an example of the event data transfer process over TCP/IP. In this example, the DAQ PC and two ROESTIs are connected with the daisy-chain.
When each ROESTI receives the trigger signal, the event data is generated on each ROESTI (Fig.~\ref{fig:egTCP2}~(a)). The event data is sent to SiTCP1 of the \textit{Network Processor} and encapsulated in the Ethernet frame. Each Ethernet frame is sent to the DAQ PC or the ROESTI1. Because the ROESTI2 sends the Ethernet frame to the next ROESTI, the destination of the Ethernet frame from the ROESTI2 is the ROESTI1.
Because of MAC address match, the Ethernet frame is sent to the \textit{SiTCP0} in the \textit{Network Processor} of the ROESTI1 (Fig.~\ref{fig:egTCP2}~(b)). The Ethernet frame is unencapsulated and the event data is stored in the FIFO of the \textit{Data Carrier}. If the ROESTI1 receives the next trigger signal, the event data is stored in the ring buffer of the \textit{Data Carrier}. Because \textit{TCP Arbiter} handles two event data, two event data never collide. The \textit{Data Carrier} arbitrates between two event data and sends the event data to the SiTCP1 as mentioned in Section~\ref{sec:datacarrier}. In the case of the example, the \textit{Data Carrier} extracts the event data of the ring buffer. The SiTCP1 makes new Ethernet frame and the event data is sent to the DAQ PC (Fig.~\ref{fig:egTCP2}~(c)). Subsequently, the \textit{Data Carrier} extracts the event data of the FIFO. The SiTCP1 makes new Ethernet frame and the event data is sent to the DAQ PC (Fig.~\ref{fig:egTCP2}~(d)). 
Similar to this example, ROESTI receives the event data from the previous ROESTI and sends the event data to either the next ROESTI or the DAQ PC. This process is repeated in all the ROESTIs until all event data reach the DAQ PC. 

The FIFO of the \textit{Data Carrier} acts as the TCP receive buffer. When the receive buffer is full, TCP temporarily suspends the data transmission process so that the buffer does not overflow. Therefore, event data are not lost. 
In the \textit{Data Carrier}, the event data which contains smaller event number has priority. Therefore, the older event data has priority. By this process, all the ROESTIs have the same priority. It never happens that only a particular ROESTI sends event data over TCP/IP.

\section{Measurement and Result}
\label{sec:Measurement}

\begin{figure}[htb!]
 \begin{center}
 \includegraphics[width=0.3\textwidth]{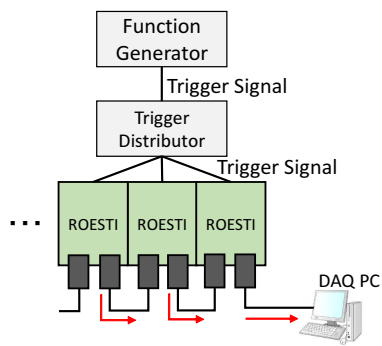}
 \end{center}
 \vspace{-5mm}
 \caption{Measurement setup.}
 \label{fig:setup}
\end{figure}

The performance of TCP data transfer was evaluated by using two to six ROESTIs. Fig.~\ref{fig:setup} shows the experimental setup. Each ROESTI and the DAQ PC were connected with the daisy-chain. The function generator generated the trigger signal periodically. The trigger signal was distributed to all the ROESTIs using the trigger distributor. After receiving the trigger signal, each ROESTI sends the one event data to the DAQ PC with the daisy-chain over TCP. The DAQ PC received the data from each ROESTI and calculated the transfer speed and throughput of the entire data. The data size of event was set to 37112 bytes, which was corresponding to the maximum event data size. The DAQ PC was a Sen-SV9R-LCi7EX-TMZ-VEditor4K, Intel Core i7-5960X (3.0-3.5 GHz, 8 Cores, 16 Threads, 20 MB Smart Cache, TDP140W) running Scientific Linux release 6.10.

\begin{figure}[htb!]
 \begin{center}
 \includegraphics[width=0.4\textwidth]{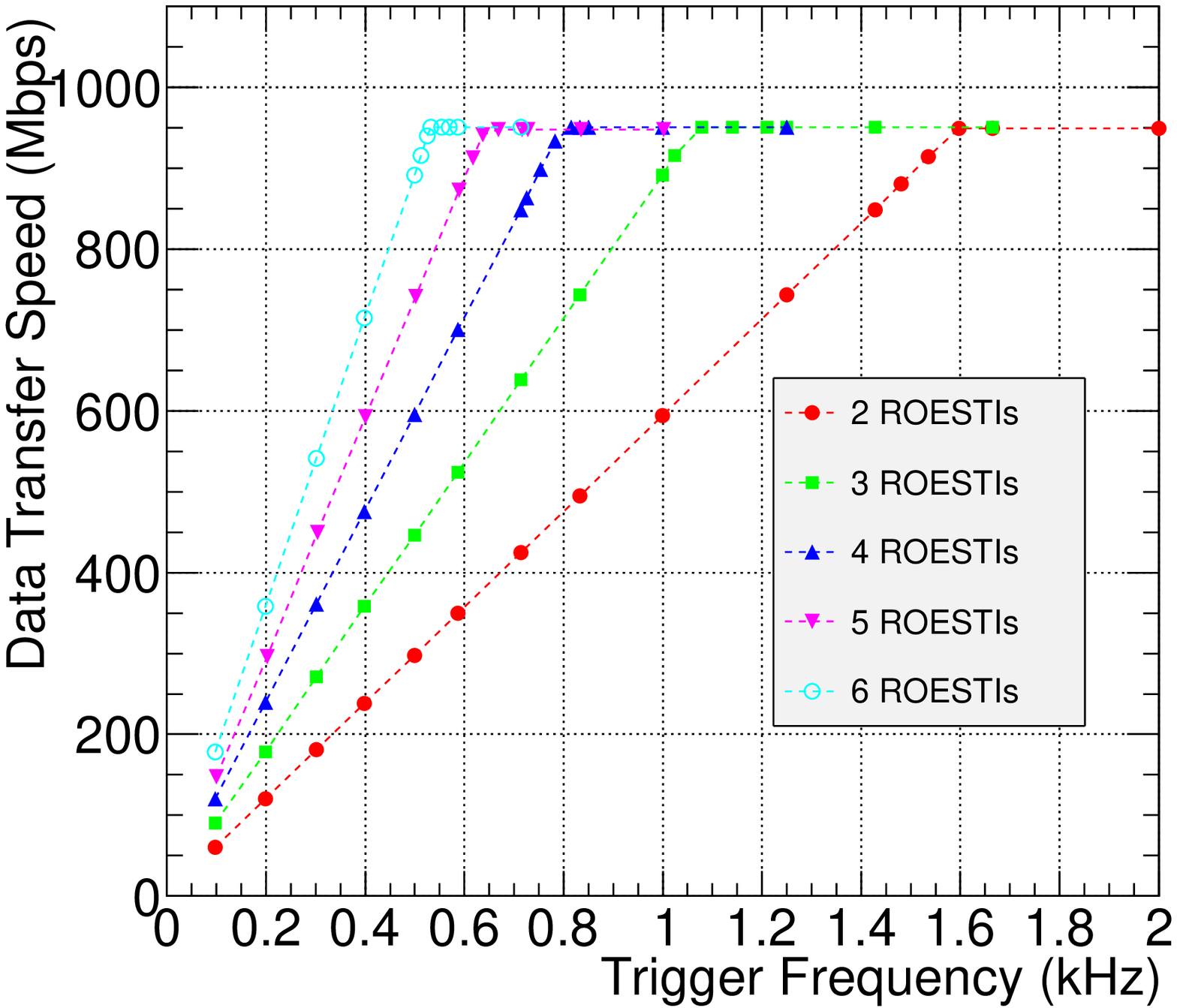}
 \end{center}
 \vspace{-5mm}
 \caption{TCP Data transfer speed as a function of trigger frequency. The difference in color indicates the difference in the number of ROESTIs connected by the daisy-chain.}
 \label{fig:freq_vs_speed}
\end{figure}

\begin{figure}[htb!]
 \begin{center}
 \includegraphics[width=0.4\textwidth]{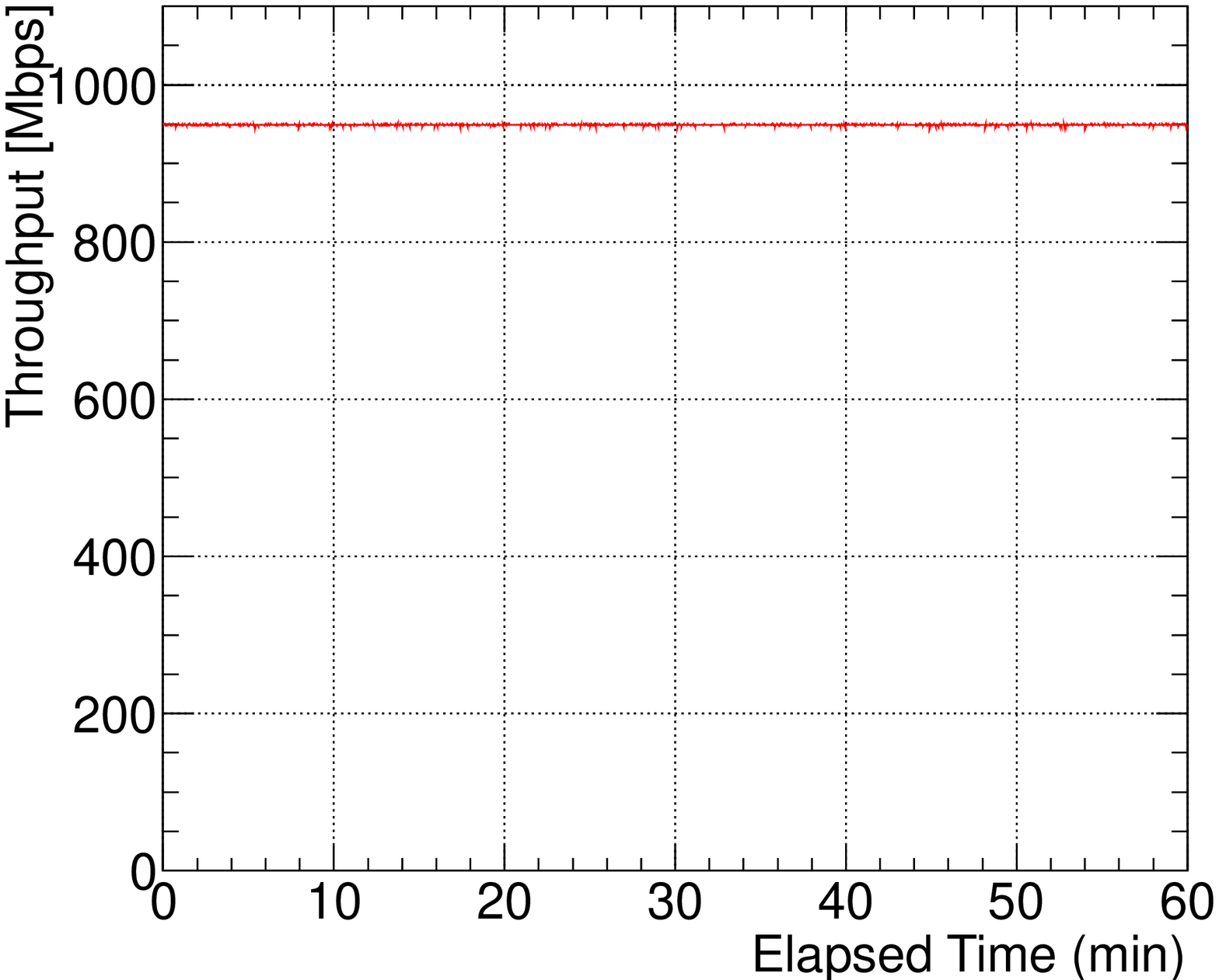}
 \end{center}
 \vspace{-5mm}
 \caption{Throughput of TCP data transfer as a function of elapsed time.}
 \label{fig:time_vs_speed}
\end{figure}

Fig.~\ref{fig:freq_vs_speed} shows the TCP data transfer speed as a function of trigger frequency with two to six ROESTIs connected by the daisy-chain. When the transfer speed was not saturated, each board sent all the event data without loss after receiving the trigger signal. Hence the data transfer speed was proportional to the trigger frequency. In any number of boards, the maximum data transfer speed was 950~Mbps, which corresponds to the theoretical limit speed of TCP over Gigabit Ethernet calculated with overheads, such as protocol headers; this implies that the throughput corresponds to the theoretical limit of TCP over Gigabit Ethernet. In the case of six boards, when the trigger frequency was 533~Hz, the transfer speed reached 950~Mbps ($= 533$~Hz~$\times$~$37112$~bytes/event~$\times$~$6$~ROESTIs) of the theoretical limit. When the trigger frequency was greater than 533~Hz, all the ROESTIs received approximately 533 trigger signals per second and other trigger signals were ignored.
Fig.~\ref{fig:time_vs_speed} shows the throughput of the TCP data transfer as a function of the elapsed time. In this measurement, six ROESTIs and the DAQ PC were connected with the daisy-chain and the trigger frequency was set at 570~Hz. To evaluate the stability, we measured the throughput of the TCP data transfer every time 1000 event data were sent to the DAQ PC. Throughput of the TCP data transfer constantly corresponded with the theoretical limit speed of TCP over Gigabit Ethernet. We also measured data loss rate when total data rate was less than the theoretical limit speed. In this measurement, we used six ROESTIs and total data transfer speed was set at 910~Mbps. All ROESTI constantly sent 100\% of the data for 1 hour. We confirmed that our Gigabit Ethernet daisy-chain function satisfied the target performance. 
In the realistic COMET situation,
five or more ROESTIs are planned to be connected with one daisy-chain line in the COMET experiment. All the ROESTIs receive the same trigger signal. After receiving the trigger signals, each ROESTI generates one event data. Average trigger rate was estimated to be 1~kHz by the simulation study. Event data size is variable and the average data size is estimated to be 10300~bytes. Total average data rate in one daisy-chain with five boards is estimated to be 410~Mbps  ($= 10300$~bytes~$\times$~$1$~kHz~$\times$~$5$~ROESTIs).
This data rate is small enough to be compared to the bandwidth of Gigabit Ethernet. Even if we connect more ROESTIs, network communication with daisy-chain is possible. 
Therefore, we are going to adopt this function in the COMET experiment.


\section{Summary}
\label{sec:Summary}

We are promoting the preparation for the COMET experiment at J-PARC, which aims to search for the neutrinoless transition of a muon to an electron. We plan to adopt the straw tube tracker as a $\mu$--$e$ conversion detector to achieve an excellent momentum resolution of better than 200~keV/c. We have developed the readout electronics board called ROESTI for the detector. To prevent the degradation of the detector signal, we plan to install the ROESTI in the gas manifold of the detector. The number of vacuum feedthroughs needs to be reduced due to space constraints and cost limitations. To decrease the number of vacuum feedthroughs drastically, we developed a new network communication scheme with the daisy-chain over Gigabit Ethernet. 
The required function for communication with the daisy-chain is to transfer data with TCP/IP over Ethernet. Additionally, a slow control function is required for setting and reading parameters of the FPGA and IC chips on each ROESTI. We implemented two SiTCPs and new circuits for the daisy-chain on the FPGA of ROESTI. 
The FPGA on ROESTIs verifies the MAC address of the Ethernet frames and controls the path of the ones. This circuit enables the DAQ PC to communicate with the target ROESTI directly for the slow control over UDP/IP.
To avoid data transfer slowdown and unstable communication by TCP re-transmission, a ROESTI sends the event data to the next ROESTI. 
This process is repeated until all the event data are finally sent to the DAQ PC. 
We measured the throughput, the stability, and the data loss rate of the Gigabit Ethernet daisy-chain function and obtained reasonable results in any number of ROESTIs.
We are going to adopt this function in the COMET experiment.



\end{document}